**Anisotropy of Magnetic Field Spectra at Kinetic Scales of Solar Wind Turbulence as Revealed by Parker Solar Probe in the Inner Heliosphere**

S. Y. Huang[1,2,♣], S. B. Xu[1], J. Zhang[1], F. Sahraoui[3], N. Andrés[4,5], J. S. He[6], Z. G. Yuan[1], X. H. Deng[7], K. Jiang[1], Y. Y. Wei[1], Q. Y. Xiong[1], Z. Wang[1], L. Yu[1], and R. T. Lin[1]

[1]School of Electronic Information, Wuhan University, Wuhan, 430072, China

[2]Hubei Luojia Laboratory, Wuhan, 430079, China

[3]Laboratoire de Physique des Plasmas, CNRS-Ecole Polytechnique-Sorbonne Université-Paris-Saclay-Observatoire de Paris-Meudon, Palaiseau, F-91128, France

[4]Departamento de Física, Facultad de Ciencias Exactas y Naturales, UBA, Ciudad Universitaria, 1428, Buenos Aires, Argentina

[5]Instituto de Astronomía y Física del Espacio, CONICET-UBA, Ciudad Universitaria, 1428, Buenos Aires, Argentina

[6]School of Earth and Space Sciences, Peking University, Beijing, 100871, China

[7]Insititute of Space Science and Technology, Nanchang University, Nanchang, 330031, China

[♣]Corresponding author: shiyonghuang@whu.edu.cn

**Abstract**

Using the Parker Solar Probe data taken in the inner heliosphere, we investigate the power and spatial anisotropy of magnetic-field spectra at kinetic scales (i.e., around sub-ion scales) in solar wind turbulence in the inner heliosphere. We find that strong anisotropy of magnetic spectra occurs at kinetic scales with the strongest power in the perpendicular direction with respect to the local magnetic field (forming an angle $\theta_B$ with the mean flow velocity). The spectral index of magnetic spectra varies from −3.2 to −5.8 when the angle $\theta_B$ changes from 90° to 180° (or 0°), indicating that strong anisotropy of the spectral indices occurs at kinetic scales in the solar wind turbulence. Using a diagnosis based on the magnetic helicity, we show that the anisotropy of the spectral indices can be explained by the nature of the plasma modes that carry the cascade at kinetic scales. We discuss our findings in light of existing theories and current development in the field.

**Introduction**

Turbulence is ubiquitous in the astrophysical and space plasmas, and plays an essential role in particle heating, energy transport and dissipation (e.g., Bruno & Carbone, 2005; Sahraoui et al. 2009, 2010, 2020; Andrés et al. 2019, 2020, 2021; Huang et al. 2010, 2012, 2014, 2017, 2020a, 2020b, 2020c, 2021a; Huang & Sahraoui, 2019). In the solar wind at the magnetohydrodynamic (MHD) scales (the range between the outer scale and the ion characteristic scale) turbulence is thought to be dominated by the cascade of Alfvénic fluctuations (Iroshnikov 1963; Kraichnan 1965; Boldyrev, 2006). Most recently, based on NI MHD theory, Zank et al (2021) argued that the MHD scale turbulence is dominated by the dissipation of quasi-2D turbulence, which can be the 2D magnetic islands or vortex structures that are commonly observed in the solar wind. In this range of the scales, the magnetic field spectra have an index around -5/3, consistent with Kolmogorov's theory of turbulence (Kolmogorov 1941; Frisch 1995). At higher temporal frequencies (i.e., below ion spatial scales), the magnetic spectra steepen and have an index varying between −2.8 to -4.5 (e.g., Goldstein et al. 1995; Leamon et al. 1998; Sahraoui et al. 2013). This range of turbulence is often referred to as a dissipation [Goldstein et al. 1995] or a dispersive range (Saito et al. 2010). Sahraoui et al. (2010) used the Cluster Search-coil data that extends the spectra to high frequencies and introduced the notion of the *transition range* (~[0.4, 4] Hz where the spectra steepen to ~−4.0 before flattening to ~ -2.8 at higher frequencies (e.g., Sahraoui et al. 2009, 2010, 2013; Huang et al. 2020b, 2021a).

In solar wind turbulence, spatial anisotropy has been studied for a long time, yet, it is still hotly debated (e.g., Horbury et al. 2008; Chen et al. 2010; Oughton et al. 2015; Horbury et al. 2012; Wang et al. 2016; Duan et al. 2021; Zhang et al. 2022). At MHD scales, Horbury et al. (2008) used 30 days of Ulysses data to compute then fit the magnetic spectra for different angles $\theta_B$ (the angle between the radial direction and the local mean magnetic field), which is used as a proxy for investigating spatial anisotropy based on the Taylor hypothesis. They found that the spectra steepen from −5/3 to −2 as $\theta_B$ gets varies from 90° to 0°. This result was later confirmed by other studies (Podesta 2009; Luo & Wu 2010), and used as an argument in favor of the critical balance (CB) conjecture (Goldreich and Sridhar, 1995), which predicts a spectrum that scales as $k_\perp^{-5/3}$ (and $k_\parallel^{-2}$) [Scheckochihin et al. 2009]. However, based on the same Ulysses's data, Wang et al. (2016) adopted a stringent sampling criterion to select $\theta_B$ and found that the spatial anisotropy becomes weaker or even vanishes, which questions the validity of the CB conjecture to solar wind turbulence. Similar conclusions were reached recently by Telloni et al. (2019) based on WIND data and Zhao et al. (2020a) by PSP's observations. Zank et al. (2020) suggested that this deviation from CB conjecture can be well explained by nearly incompressible (NI) MHD turbulence. Recently, utilizing data from PSP's first seven orbits, Zhao et al. (2022) reported the inertial-range magnetic-fluctuation anisotropy. They demonstrated that the wave-vector anisotropy is generally consistent with the two-dimensional (2D) plus slab turbulence model, and they further determined the fraction of power in the 2D versus slab component.

At kinetic scales, Chen et al. (2010) used a multi-spacecraft method to estimate the second-order structure functions $S_\parallel$ and $S_\perp$ of magnetic fluctuations at different angles $\theta_B$. They showed that the perpendicular component is the major population, implying anisotropic fluctuations with $k_\perp > k_\parallel$. The spectra of the perpendicular and parallel component have slightly different indices at different field-separation angles. Recently, Huang et al. (2021b) have found two populations in the two-dimensional spatial correlation functions (SCFs) in the kinetic range of magnetosheath turbulence: the minor component elongated along the perpendicular direction while the major one was elongated along the parallel direction, indicating that the distribution of magnetosheath turbulence in the wavenumber space is dominantly transverse to the background magnetic field with a weak component along the magnetic field (Sahraoui et al. 2006). Wang et al. (2020) calculated five-point second-order structure functions of magnetic field in the magnetosheath to investigate three-dimensional anisotropies at sub-ion scales, and found that the ratio between statistical eddies' parallel and perpendicular lengths features a trend of rise then fall whereas the anisotropy in the perpendicular plane appears scale-invariant. Most recently, Duan et al. (2021) reported anisotropic magnetic energy spectra at kinetic scales in inner heliosphere, where the spectral index varies from −3.7 to −5.7 in the transition range and −2.57 to −2.9 in the ion-electron scales when the angle $\theta_{VB}$ between the plasma flow and the magnetic field changes from 90° to 180°. Note that the CB conjecture when extended to sub-ion scales predicts a spectrum that is much steeper in the parallel direction (and $k_\parallel^{-5}$) then in the perpendicular one ($k_\perp^{-7/3}$) (Schekochihin et al. 2009; Cho & Lazarian, 2004). However, such an anisotropy of the magnetic field spectra at kinetic scales has not been verified in the solar wind.

In the present study, the high-time resolution data from the Parker Solar Probe (PSP) mission located in the inner heliosphere (about 0.17 astronomical unit, AU) are used to investigate the anisotropy of magnetic field spectra at kinetic scales for different types of the solar wind and for different angles $\theta_B$. When selecting and classifying the data samples as function of the angle $\theta_B$, both the criteria of Horbury et al. (2008) and those of Wang et al. (2016) were used separately to judge their (possible) impact on the results of the study.

**Data and Methods**

Measurements from the PSP mission's first solar encounter from October 31$^{st}$, 2018 to November 11$^{th}$, 2018 are used in this study. The FIELDS flux-gate magnetometer (FGM) provides the data of magnetic field vector (Bale et al. 2016; Malaspina et al. 2016), and the Solar Wind Electron, Alpha, Proton (SWEAP) experiment provides the proton moments (including density, velocity and temperature) data (Kasper et al. 2016; Case et al. 2020).

We utilize the Morelet wavelet to compute the PSD and the Gaussian-window method to obtain $\theta_B$ as function of time and frequency (Horbury et al. 2008; Podesta 2009; He et al. 2011). Then, the $\theta_B$ values ranging from 0° to 180° were divided into 18 bins, and the spectra were classified into 18 categories. Two methods are used to determine which and how the samples should be selected. With the first method (Method 1) adopted by Horbury et al. (2008), the spectra are simply divided into 18 categories according to the value of $\theta_B$. For each frequency ($f$) and each angle bin ($i$), the average PSD is calculated by

$$\text{Method 1:} \quad PSD(i, f) = \frac{1}{N(i, f)} \times \sum_{\theta_B(t,f) \in [10(i-1), 10i]} PSD(t, f)$$

in which $i$ ranges from 1 to 18, and $N(i, f)$ is the total number of samples in category $i$ for each frequency. The second method (Method 2), proposed by Wang et al. (2016), sets a stringent criterion to select the spectral samples. According to Podesta (2009), in order to calculate PSD at time $t_k$ for frequency $f$, the magnetic field data in the time interval of $[t_k - 1.5\tau, t_k + 1.5\tau]$ (where

$\tau \sim 1/f$) are mainly used in the wavelet method. Therefore, if all $\theta_B$ in this time interval belongs to the same angle bin, then the spectra at the time point $t_k$ is retained within the corresponding category. Wang et al. (2016) selected the wavelet coefficients when the angles $\theta_B$ at the time instants $t_k - 1.5\tau$, $t_k$, and $t_k + 1.5\tau$ are all in this bin. In this way, for each frequency $f$ and each angle bin, the averaged PSD reads is

$$\text{Method 2:} \quad PSD(i,f) = \frac{1}{N(i,f)} \times \sum_{\substack{\theta_B(t,f) \in [10(i-1),10i] \\ \theta_B(t-1.5\tau,f) \in [10(i-1),10i] \\ \theta_B(t+1.5\tau,f) \in [10(i-1),10i]}} PSD(t,f).$$

Furthermore, the normalized reduced magnetic helicity ($\sigma_m$) is calculated using a wavelet transform (He et al. 2011; Huang et al. 2020b). The $\sigma_m$ can be used to diagnose the polarization of solar wind turbulence (Matthaeus & Goldstein 1982), which can be linked to the classical wave polarization (see, e.g., Howes & Quataert 2010; Meyrand & Galtier 2012; Klein et al. 2014; He et al. 2011, 2012; Huang et al. 2020b). Here, we also use the two methods described above to get the averaged $\sigma_m$ for each $\theta_B$ bin and each frequency $f$, namely

$$\text{Method 1:} \quad \sigma_m(i,f) = \frac{1}{N(i,f)} \times \sum_{\theta_B(t,f) \in [10(i-1),10i]} \sigma_m(t,f),$$

$$\text{Method 2:} \quad \sigma_m(i,f) = \frac{1}{N(i,f)} \times \sum_{\substack{\theta_B(t,f) \in [10(i-1),10i] \\ \theta_B(t-1.5\tau,f) \in [10(i-1),10i] \\ \theta_B(t+1.5\tau,f) \in [10(i-1),10i]}} \sigma_m(t,f).$$

**Results**

Figure 1 shows the magnetic spectra for different $\theta_B$ bins obtained using Method 1 (a) and Method 2 (b). To separate the spectra of different angle bins, the values of the PSD shown in Figure 1 result from multiplying different coefficients of different $\theta_B$ bins. The frequency band ranges from 0.008 to 10 Hz, which includes the MHD inertial range (from 0.01 Hz to 1 Hz) and the kinetic scales (from 2 Hz to ~10 Hz). We fit magnetic spectra as a function of frequency (i.e., $f^\alpha$) at the MHD scales, while the spectral indices are shown in Figure 1c and 1d, respectively. Some indices at small $\theta_B$ are not available because there were a few spectra samples. The spectral indices range from −1.48 to −1.62 for Method 1 and from −1.45 to −1.63 for Method 2, indicating that there is no apparent anisotropy at MHD scales in the solar wind, which may be the nature of the solar wind turbulence during the time interval of our interest.

Figure 2a-2b show the joint distributions of normalized reduced magnetic helicity $\sigma_m(\theta_B, f)$ as a function of $\theta_B$ and frequency $f$ based on Method 1 and Method 2, respectively. The normalized reduced magnetic helicity can be divided into two populations at kinetic scales: the first population has negative values of $\sigma_m$ for frequencies between 2 Hz and 10 Hz and angles $40° < \theta_B < 140°$, the second population has positive values of $\sigma_m$ for frequencies between 2 Hz and 8 Hz and angles $\theta_B < 30°$ or $\theta_B > 150°$. For an inward-oriented background magnetic field ($B_R < 0$), a forward left-handed polarized wave mode has positive magnetic helicity, while a forward right-handed polarized wave mode has negative magnetic helicity (e.g., He et al. 2011). The magnetic fluctuations with $\theta_B$ close to 0° or 180° correspond to waves propagating quasi-parallel or quasi-antiparallel to the mean magnetic field, while the magnetic fluctuations with the $\theta_B$ close to 90° correspond to waves propagating quasi-perpendicular to the mean magnetic field (e.g., He et al. 2011, 2015). Therefore, the magnetic fluctuations with positive $\sigma_m$ around 0° or 180° can be identified as quasi-parallel left-handed Alfvén ion Cyclotron Waves (ACWs), while the magnetic fluctuations with negative helicity around 90° are likely to be quasi-perpendicular right-handed kinetic Alfvén waves (KAWs, He et al. 2011; Bruno & Telloni 2015; Huang et al. 2020b; Zhao et al., 2021a). It should be noted that positive $\sigma_m$ around 0° or 180° can also correspond to inward fast/whistler waves (Zhao et al., 2021b). However, it is hard to exclude the contribution of inward fast/whistler waves in our time interval indeed because it requires high-time resolution (< ~0.1 s at least) plasma data which can't be provided by PSP (Zhao et al., 2020b).

The spectra at kinetic scales (integrated in frequency) for different $\theta_B$ are shown in Figure 2c-2d. One can see that the integrated power is higher in the quasi-perpendicular direction ($\theta_B \sim 90°$), and tends to decrease when the $\theta_B$ turns to the parallel (and anti-parallel) direction. Figure 2e-2f show the angular variation of spectral indices (black curve) and magnetic helicity $\sigma_m$ (red curve), calculated by their averaged values within kinetic scales. The spectra have a slope around −3.4 in the perpendicular direction, and becomes (continuously) steeper when $\theta_B$ turns to the (anti-) parallel direction. The steepest spectrum appears in the bin 170° <$\theta_B$<180°, with a slope −5.60 ± 0.18 (Method 1) and −5.76 ± 0.14 (Method 2). Moreover, compared with the trend of the spectra index and $\sigma_m$, one can see that the positive magnetic helicities are always accompanied by steeper spectra. In contrast, the negative magnetic helicities correspond to smaller spectral indices. This suggests a correlation between the presence of quasi-parallel left-handed (ACWs) fluctuations and the steep spectra at kinetic scales, while flatter ones seem to correlation with quasi-perpendicular right-handed (KAWs) fluctuations.

In addition, we investigate the variations of spectra index and magnetic helicity $\sigma_m$ in the fast ($V_f$ > 380 km/s, <$V_f$>= 469 km/s) and the slow solar wind ($V_f$ ≤ 380 km/s, <$V_f$ >= 295 km/s). Considering that PSP observes little solar wind with the velocity above 500 km/s and the solar wind speed increases with the radial distances during perihelion 1 of PSP (e.g., Huang et al., 2020d), hence it is credible to use the threshold of 380 km/s when close to the Sun to select fast wind in present study. Figure 3 shows the angular variations of spectral indices $\alpha$ and $\sigma_m$ for both types of winds. In the slow solar wind, $\sigma_m$ is about −0.2 in the perpendicular direction with $\alpha$ ~ −3.2 and about 0 in the antiparallel direction with $\alpha$ ~ −5.7, while $\sigma_m$ can reach −0.4 in the perpendicular direction with $\alpha$ ~ −3.3 and 0.3 at the antiparallel direction with $\alpha$ ~ −6.7 in fast winds. These results suggest that the KAWs and ACWs in the fast solar wind are more significant than the ones in the slow solar wind and the magnetic field spectra become steeper with positive $\sigma_m$ in the antiparallel direction for the fast solar wind. Moreover, the magnetic spectra of fast solar wind show much stronger degree of anisotropy, with spectral indices reaching −6.7 in the bin of 170° <$\theta_B$<180°. These results support the finding above that the steep spectra (at kinetic scales) are closely related to ACWs while flatter ones seem to be carried by KAWs fluctuations.

**Discussions and Conclusions**

In the present study, we investigated the anisotropy of magnetic-field spectra at kinetic scales in solar wind turbulence using the observations from Parker Solar Probe made in the inner heliosphere. Two different methods have been utilized to select samples of $\theta_B$. The results lead to the same conclusion: the spectra have an anisotropic power at kinetic scales with the strongest values in the perpendicular direction ($\theta_B \sim 90°$). The spectral index varies from $-3.2$ to $-5.8$ when the angle $\theta_B$ varies from $90°$ to $180°/0°$, concomitant with the variation of the magnetic helicity from negative to positive values. The observed anisotropy seems to be controlled by the nature wave modes that carry the cascade at kinetic scales: the flat spectra with $\theta_B \sim 90°$ originate from KAWs, while steep spectra with $\theta_B \sim 0°$ or $180°$ are come for ACWs. A distinction between slow and fast wind did not show a significant difference in the anisotropy of the spectral indices. However, a stronger anisotropy of the spectral indices is observed in fast wind with larger magnetic helicities, which further confirms that the nature of wave modes at kinetic scales determine the anisotropy of spectral index.

Sahraoui et al. (2010) and Chen et al. (2010) have found the magnetic power at kinetic scales are dominated by quasi-perpendicular wavenumber in the solar wind at 1 AU. Chen et al. (2010) have shown the spectral index at kinetic scales varies from $-3$ along the parallel direction to $-2.6$ along the perpendicular direction, which is much flatter than our results in the inner heliosphere. Duan et al. (2021) have reported anisotropic spectral indices that vary from $-5.7\pm1.3$ to $-3.7\pm0.3$ in the transition range (i.e., the kinetic scales defined in our work) from the parallel to the perpendicular directions with respect to the ambient magnetic field, which is compatible with our results. In addition, they found that the spectral index varies from $-2.9\pm0.2$ to $-2.57\pm0.07$ in the ion-electron range (with the frequencies higher than those of the transition range). However, they did not give the possible explanation for the reported anisotropy. Our results confirm the close relationship between the magnetic helicities and the spectral indices at kinetic scales: the shallow spectra are controlled by the KAWs, while the steep ones are determined by the ACWs. Moreover, the steep power spectral indices at kinetic scales from the parallel to perpendicular directions prove that there is a transition range signature in all directions. This also confirms the model proposed by Sahraoui et al. (2009, 2010) that the second inertial range between ion and electron scales occurs above the transition where part of the energy is damped into proton heating. Based on the polarization of the fluctuations inferred from the magnetic helicity diagnosis, our results suggest that protons are likely to be heated by the KAWs or ACWs at kinetic scales.

Most recently, Meyrand et al. (2021) proposed that "helicity barrier" near the ion scales could prevent the energy from cascading to smaller scales and give rise to a steep transition range in $\beta \ll 1$ plasma in the finite-Larmor-radius MHD turbulence. In their simulations, the perpendicular spectral indices near the ion scales can reach a value about −3.8. Using six-dimensional hybrid-kinetic simulations, Squire et al. (2021) found that the "helicity barrier" could drive ion heating by ACWs. They also observed a steep transition range with a spectral index up to −4 in the perpendicular direction accompanied by an enhanced (negative) $\sigma_m$ which corresponds to KAWs, and an even steeper parallel spectrum ($\alpha \sim -6$) with large positive $\sigma_m$ which corresponds to ACWs. Considering that Squire et al. (2021) performed their simulations with a larger $\beta$ ($\sim 0.3$) plasma, which is closer to the values observed in the inner heliosphere, the mechanism of "helicity barrier" seems to be a possible explanation of our present results.

**Acknowledgement**

This work was supported by the National Natural Science Foundation of China (42074196, 41874191, 41925018, 41874200), and the National Youth Talent Support Program. SYH acknowledges the project supported by Special Fund of Hubei Luojia laboratory. N.A. acknowledges financial support from CNRS/CONICET Laboratoire International Associé (LIA) MAGNETO and financial support from the following grants: PICT 2018 1095 and UBACyT 20020190200035BA. The datasets analyzed in present study are publicly available from the NASA's Space Physics Data Facility (SPDF) at https://spdf.gsfc.nasa.gov/pub/data/psp/.

**References**

Andrés, N., Sahraoui, F., Galtier, S., et al. 2019, Phys. Rev. Lett., 123, 245101

Andrés, N., Romanelli, N., Hadid, L.Z., Sahraoui, F., DiBraccio, G. and Halekas, J., 2020. The Astrophysical Journal, 902(2), p.134

Andrés, N., Sahraoui, F., Hadid, L.Z., Huang, S.Y., Romanelli, N., Galtier, S., Dibraccio, G. and Halekas, J., 2021. The Astrophysical Journal, 919(1), p.19

Bale, S. D., Goetz, K., Harvey, P. R., et al. 2016, SSRv, 204(1), 49-82. https://doi.org/10.1007%2Fs11214-016-0244-5

Boldyrev, S., 2006, PRL, 96. 115002. https://doi.org/10.1103/PhysRevLett.96.115002

Bruno, R., & Carbone, V. 2005, Living Rev. Sol. Phys, 2, 4. https://doi.org/10.12942/lrsp-2005-4

Bruno, R., & Telloni, D. 2015, ApJL, 811, L17. https://doi.org/10.1088/2041-8205/811/2/L17

Case, A. W., Kasper, J. C., Stevens, M. L., et al. 2020, ApJS, 246(2), 43. https://doi.org/10.3847/1538-4365/ab5a7b

Chen, C. H. K., Horbury, T. S., Schekochihin, A. A., et al. (2010). PhRvL, 104(25), 255002. http://doi.org/10.1103/PhysRevLett.104.255002

Cho, J., & Lazarian, A. 2004, ApJ, 615, L41. http://dx.doi.org/10.1086/425215

Duan, D., He, J., Bowen, T. A., et al. 2021, ApJL, 915(1), L8. https://doi.org/10.3847/2041-8213/ac07ac

Frisch, U. 1995, Turbulence; The Legacy of A.N. Kolmogorov (Cambridge: Cambridge Univ. Press)

Goldreich, P., & Sridhar, S. 1995, ApJ, 438, 763. http://dx.doi.org/10.1086/175121

Leamon, R. J., Matthaeus, W. H., Smith, C. W., & Wong, H. K. 1998, JGR, 103, 4775. http://adsabs.harvard.edu/abs/1998JGR...103.4775L

He, J., Marsch, E., Tu, C., Yao, S., & Tian, H. 2011, ApJ, 731(2), 85. http://doi.org/10.1088/0004-637X/731/2/85;

He, J., Tu, C., Marsch, E., & Yao, S. 2012, ApJL, 745(1), L8. http://doi.org/10.1088/2041-8205/745/1/L8

Horbury, T. S., Forman, M., & Oughton, S. 2008, Phys. Rev. Lett., 101, 175005. http://doi.org/10.1103/PhysRevLett.101.175005

Howes, G. G., Cowley, S. C., Dorland, W., Hammett, G. W., Quataert, E., & Schekochihin, A. A. 2008, JGRA, 113, 5103. http://adsabs.harvard.edu/abs/2010ApJ...709L..49H

Horbury, T., Wicks, R., & Chen, C. 2012, Space Science Reviews, 172, 325

Huang, J., Kasper, J. C., Vech, D., et al. 2020d, ApJS, 246(2), 70. https://doi.org/10.3847/1538-4365/ab74e0

Huang, S. Y., M. Zhou, F. Sahraoui, et al. 2010, JGRA, 115, A12211, https://doi.org/10.1029/2010JA015335.

Huang, S. Y., Zhou, M., Sahraoui, F., et al. 2012. GeoRL, 39(11). https://doi.org/10.1029/2012GL052210

Huang, S., Sahraoui, F., Deng, X., et al. 2014, ApJL, 789(2), L28. http://doi.org/10.1088/2041-8205/789/2/L28

Huang, S. Y., L. Z. Hadid, F. Sahraoui, Z. G. Yuan, & X. H. Deng. 2017, APJL, 836, L10, http://doi.org/10.3847/2041-8213/836/1/L10

Huang, S. Y., and F. Sahraoui (2019), Astrophys. J. 876, 138

Huang, S. Y., Zhang, J., Sahraoui, F., et al. 2020a, ApJL, 898(1), L18. https://doi.org/10.3847/2041-8213/aba263

Huang, S. Y., Wang, Q. Y., Sahraoui, F., et al. 2020c, ApJ, 891(2), 159. https://doi.org/10.3847/1538-4357/ab7349

Huang, S. Y., Zhang, J., Sahraoui, F., et al. 2020b, ApJL, 897(1), L3. https://doi.org/10.3847/2041-8213/ab9abb

Huang, S. Y., Sahraoui, F., Andrés, N., et al. 2021a, ApJL, 909(1), L7. https://doi.org/10.3847/2041-8213/abdaaf

Huang, S. Y., Xiong, Q. Y., Yuan, Z. G., et al. 2021b, JGRA, 126, e2020JA028780. https://doi.org/10.1029/2020JA028780

Iroshnikov, P. S. 1963, AZh, 40, 742. https://ui.adsabs.harvard.edu/#abs/1964SvA.....7..566I/abstract

Kasper, J. C., Abiad, R., Austin, G., et al. 2016, SSRv, 204(1), 131-186. https://doi.org/10.1007/s11214-015-0206-3

Klein, K. G., Howes, G. G., & TenBarge, J. M. 2014, ApJL, 790, L20. https://doi.org/10.1088/2041-8205/790/2/L20

Kraichnan, R. H. 1965, PhFl, 8(7), 1385-1387. https://doi.org/10.1063/1.1761412

Kolmogorov, A. 1941, DoSSR, 30, 301. https://ui.adsabs.harvard.edu/abs/1941DoSSR..30..301K/abstract.

Luo, Q. Y., & Wu, D. J. 2010, ApJL, 714, L138. http://adsabs.harvard.edu/abs/2010ApJ...714L.138L

Malaspina, D. M., Ergun, R. E., Bolton, M., et al. 2016, JGRA, 121, 5088, https://doi.org/10.1002/2016JA022344

Matthaeus, W. H., & Goldstein, M. L. 1982, J. Geophys. Res., 87, 6011. http://dx.doi.org/10.1029/JA087iA08p06011

Meyrand, R., & Galtier, S. 2012, PhRvL, 109, 194501. http://adsabs.harvard.edu/abs/2012PhRvL.109s4501M

Meyrand, R., Squire, J., Schekochihin, A. A., & Dorland, W. 2021, JPlPh, 87(3). https://doi.org/10.1017/S0022377821000489

Oughton, S., Matthaeus, W., Wan, M., & Osman, K. 2015, Philosophical Transactions of the Royal Society A: Mathematical, Physical and Engineering Sciences, 373, 20140152

Podesta, J. J. 2009, ApJ, 698, 986. http://adsabs.harvard.edu/abs/2009ApJ...698..986P

Sahraoui, F., Goldstein, M. L., Robert, P., & Khotyaintsev, Y. V. 2009, PhRvL, 102(23), 231102. https://doi.org/10.1103/PhysRevLett.102.231102

Sahraoui, F., Goldstein, M. L., Belmont, G., Canu, P., & Rezeau, L. 2010, PhRvL, 105(13), 131101. https://doi.org/10.1103/PhysRevLett.105.131101

Sahraoui, F., Huang, S. Y., Belmont, G., et al. 2013, ApJ, 777(1), 15. http://doi.org/10.1088/0004-637X/777/1/15

Sahraoui, F., Hadid, L. Z., & Huang, S. Y. 2020, RvMPP, 4, 1

Saito, S., Peter Gary, S., & Narita, Y. 2010, PhPl, 17, 122316

Schekochihin, A. A., Cowley, S. C., Dorland, W., et al. 2009, ApJS, 182, 310. http://dx.doi.org/10.1088/0067-0049/182/1/310

Squire, J., Meyrand, R., Kunz, M. W., et al. 2021. arXiv preprint arXiv:2109.03255.

Telloni, D., Carbone, F., Bruno, R., et al. 2019, ApJ, 887, 160

Wang, T., He, J., Alexandrova, O., Dunlop, M., & Perrone, D. 2020, ApJ, 898(1), 91. https://doi.org/10.3847/1538-4357/ab99ca

Wang, X., Tu, C., Marsch, E., He, J., & Wang, L. 2016, ApJ, 816, 15. http://adsabs.harvard.edu/abs/2016ApJ...816...15W

Zank, G. P., Nakanotani, M., Zhao, L. L., Adhikari, L., & Telloni, D. 2020, ApJ, 900(2), 115. https://doi.org/10.3847/1538-4357/abad30

Zank, G. P., Zhao, L. L., Adhikari, L., et al. 2021, PhPl, 28(8), 080501. https://doi.org/10.1063/5.0055692

Zhao, L. L., Zank, G. P., Adhikari, L., et al. 2020a, ApJ, 898(2), 113. https://doi.org/10.3847/1538-4357/ab9b7e

Zhao, L. L., Zank, G. P., Adhikari, L., & Nakanotani, M. 2022, ApJL, *924*(1), L5.

https://doi.org/10.3847/2041-8213/ac4415

Zhao, L. L., Zank, G. P., He, J. S., et al. 2021a, A&A, 656, A3. https://doi.org/10.1051/0004-6361/202140450

Zhao, L. L., Zank, G. P., He, J. S., et al. 2021b, ApJ, 922(2), 188. https://doi.org/10.3847/1538-4357/ac28fb

Zhao, J., T. Wang, D. B. Graham, J. He, W. Liu, W. M. Dunlop, and D. Wu, 2020b, ApJ, 890, 17

**Figure captions**

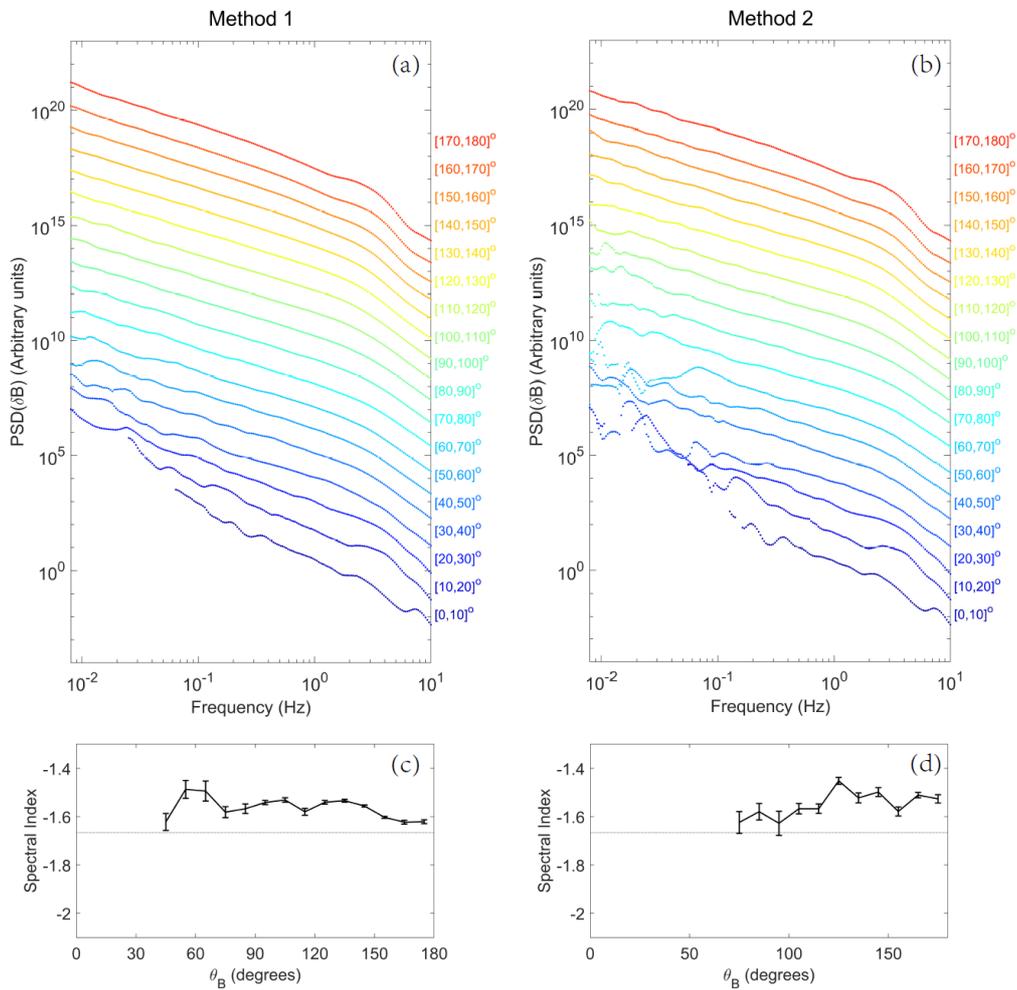

**Figure 1.** Magnetic spectra and spectral indices: magnetic spectra at 18 angular bins based on Method 1 (a) and Method 2 (b), where the PSDs are shifted in the Y-axis for clarity; spectral indices at MHD scales as a function of $\theta_B$ using Method 1 (c) and Method 2 (d).

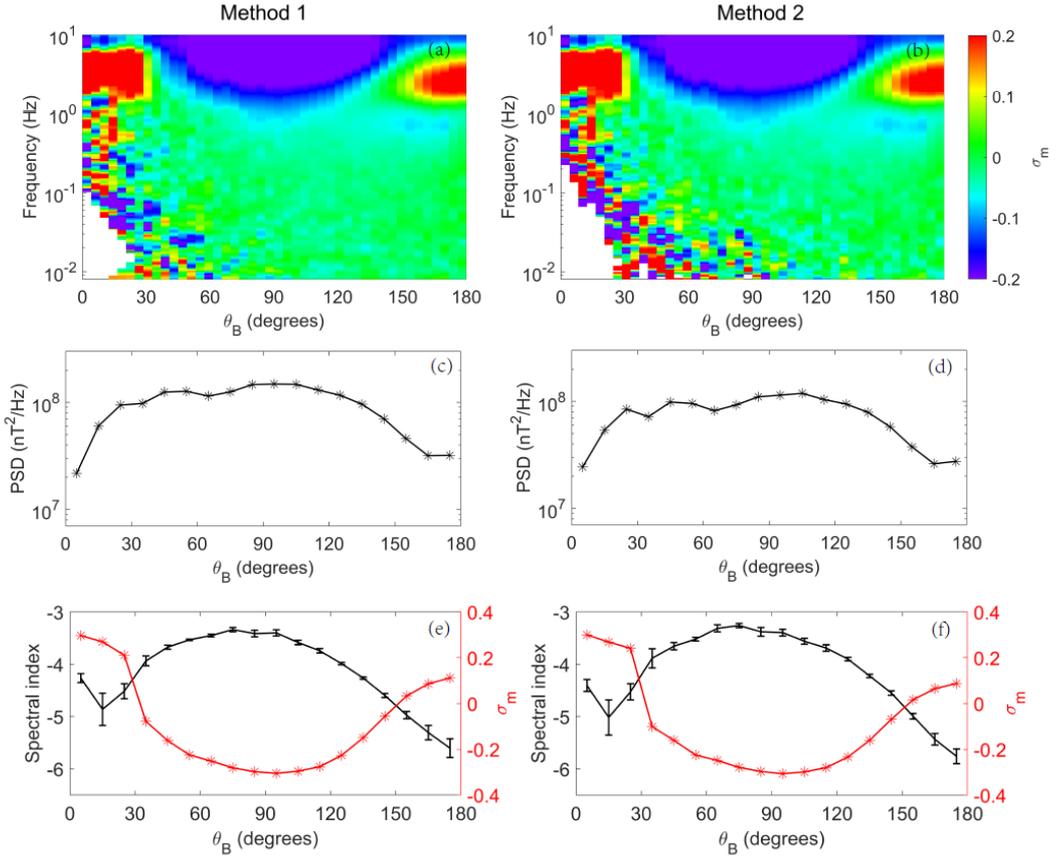

**Figure 2.** (a, b): The joint distributions obtained using the two methods of normalized reduced magnetic helicity σ$_m$ as a function of frequency and $\theta_B$. (c, d): The averaged power of spectral densities at kinetic scales in different angular bins. (e, f): The angular variations of the spectral index (black line curves) and magnetic helicity σ$_m$ (red curves) at kinetic scales.

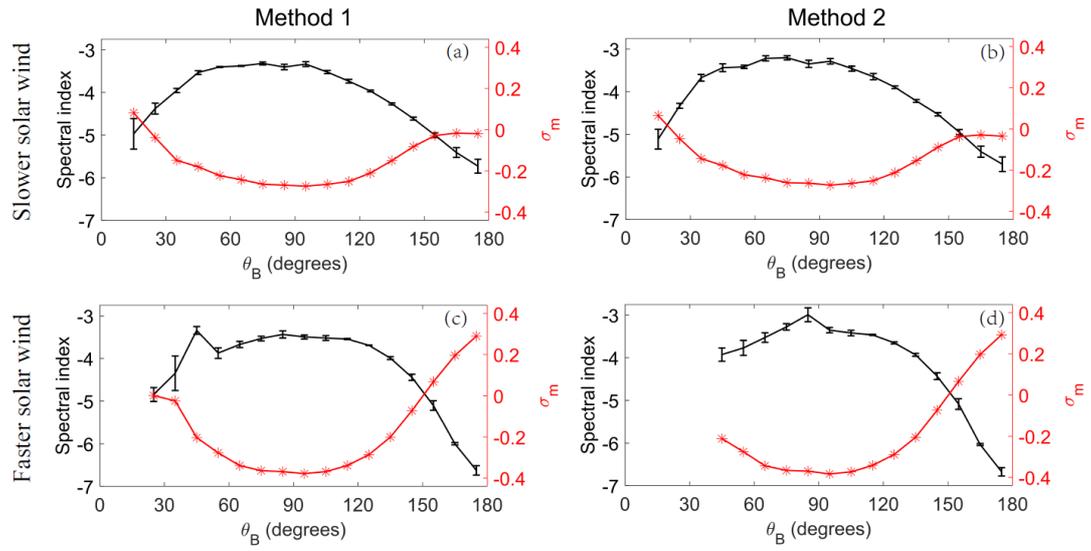

**Figure 3.** The spectral index (black curves) and magnetic helicity σ$_m$ (red curves) at kinetic scales in the slow solar wind (a, b) and the fast solar wind (c, d) based on two selection methods.